\documentclass[aps,prl,amsmath,twocolumn,floats,floatfix,superscriptaddress,nofootinbib,showpacs]{revtex4-1}

\usepackage{amsfonts,amsmath,units,wasysym,epsfig,graphicx,verbatim,color,subfigure,graphicx}
\usepackage{amsmath}
\usepackage{amssymb}
\usepackage{amsfonts}
\usepackage{hyperref}
\usepackage{bm}
\usepackage{color}
\usepackage[normalem]{ulem}
\usepackage{natbib}
\usepackage{graphicx}
\usepackage{xcolor}

\def\be{\begin{equation}}
\def\ee{\end{equation}}
\def\bea{\begin{eqnarray}}
\def\eea{\end{eqnarray}}

\begin{document}
\bibliographystyle{unsrt}


\newcommand{\rhat}{\hat{r}}
\newcommand{\iotahat}{\hat{\iota}}
\newcommand{\phihat}{\hat{\phi}}
\newcommand{\h}{\mathfrak{h}}
\newcommand{\vek}[1]{\boldsymbol{#1}}
\newcommand{\IUCAA}{\affiliation{Inter-University Centre for Astronomy and Astrophysics, Post Bag 4, Ganeshkhind, Pune 411 007, India}}
\newcommand{\WSU}{\affiliation{Department of Physics \& Astronomy, Washington State University, 1245 Webster, Pullman, WA 99164-2814, U.S.A}}

\title{Generalized source multipole moments of dynamical horizons in binary black hole mergers}
    
\author{Vaishak Prasad}
\affiliation{Inter-University Centre for Astronomy and Astrophysics, Post Bag 4, Ganeshkhind, Pune 411 007, India}

\date{\today}

\begin{abstract}
In this work,  we uncover new features in the evolution of the deformations of the dynamical horizon geometry in a binary black hole merger scenario using numerical relativity. First, in the inspiral phase, owing to the deformations, the dynamical horizons of the two black holes are found to steadily acquire multipole moments that vanish when the horizons are isolated. Out of these, the dominant moment is found to be the quadrupole moment. Second, we show that they encode detailed information about the dynamics of the binary black hole system. The dominant quadrupole multipole moment is particularly shown to be strongly correlated with the gravitational field of the system at future null infinity. Therefore, the gravitational waves carried away from the system contain information about the geometrical structure of the black holes in the strong-field regime. Third, we also find that, in the post-merger phase, the multipolar structure of the outer common dynamical horizon of the system is strongly correlated with that of the individual horizons just before the merger. The outer common horizon then settles down to equilibrium as suggested by the decay of the multipole moments gained by the system through the inspiral phase.
\end{abstract}

\maketitle

\noindent \emph{Introduction:} Numerous binary black hole mergers have been observed to date starting with the first detection in 2015 \cite{Abbott:2016blz}, 
\cite{LIGOScientific:2018mvr,TheLIGOScientific:2016pea,Nitz:2018imz,Nitz:2019hdf,Venumadhav:2019lyq,Zackay:2019tzo}, \cite{theligoscientificcollaboration2021gwtc21}, \cite{theligoscientificcollaboration2021search}.
The parameters of these binary systems,
including the masses and spins of the individual black holes, can be
inferred from the observed data \cite{TheLIGOScientific:2016wfe}. These observations have been used to extract information about the overall dynamics of the binary black hole system. The observations have so far been found to be consistent with standard general relativity \cite{TheLIGOScientific:2016src,LIGOScientific:2019fpa,Abbott:2018lct}. 

The dynamics of gravitational radiation in the far field regime, far away from the horizons of the merging black holes is fairly well understood. In particular, the multipole moments of gravitational radiation are relevant here \cite{thorne1980, Bondi:1962px, Sachs:1962wk, Newman:1962, janais:1965}. In the strong field regime, the physical boundaries of the black hole regions are described by their dynamical horizons, which are located inside the event horizons and are outside of the domain of outer communications. In recent times, advances in numerical relativity have paved way for understanding the dynamics of the binary black hole system in the strong field regime, particularly of the dynamical horizons of black holes. There have been efforts in the recent past directed towards understanding of the dynamics of the horizons, but mostly of axisymmetric scenarios like head-on collisions \cite{Owen:2009sb, Pook-Kolb:2020zhm, Pook-Kolb:2020jlr}. However understanding of the dynamics of the horizon geometry in the inspiral, merger and ringdown phase of generic and realistic non-axisymmetric dynamical scenarios, like that of a binary black hole merger, is still in its infancy \cite{Gupta:2018znn}.

In binary black hole systems, the horizons of the black holes exist in a tidal environment, possess an influx of of energy and momentum and are hence dynamical. In such a scenario, in this work we find that the general features of the dynamics of the black holes such as their orbital phasing, relative location of the black holes, etc are imprinted on the geometry of their dynamical horizons. In our previous works, the correlations between the gravitational fields at two surfaces: the dynamical horizon and future null infinity were studied. In particular, the infalling radiation at the dynamical horizons was found to be strongly correlated with the news function of the outgoing  gravitational radiation from the system \cite{shear-news2020}. In another recent study, the axisymmetric deformations of the dynamical horizons in binary black hole scenario were studied and were used to quantify the strong field tidal deformability of black holes \cite{prasad2021tidal}. In this work, we also add to the understanding of the strong field dynamics of black holes in general relativity (i.e. at the dynamical horizons) and its correlation with the weak field dynamics (i.e. at future null infinity) by studying the evolution of the horizon geometry of dynamical horizons, described by its source multipole moments, for the first time in a binary black hole scenario through the inspiral, merger and ringdown phases. We then show that not just the infalling radiation at the horizon, but the geometry of the gravitational field in the strong field regime i.e. that of the dynamical horizon itself is strongly correlated with the spacetime geometry at null infinity. Thus, although future directed causal curves from the dynamical horizons cannot reach $\mathcal{I}^+$, one can potentially use these results to discern the multipolar structure of the deformed horizon geometries using information at $\mathcal{I}^{+}$.  We also find that the evolution of the multipole moments of the system can be described by a generic expansion in terms of the distance of separation between the black holes. This was previously reported in \cite{Cabero:2014nza, prasad2021tidal} and here, we extend it to more general non-axisymmetric deformations of the dynamical horizons.

\noindent \emph{Basic notions:} The main quantities of interest are (a). the generalized source multipole moments of the dynamical horizons, which are defined at a dynamical horizon $\mathcal{H}$, \cite{Ashtekar:2004cn,Booth:2005qc} obtained by a time evolution of marginally trapped surfaces, and (b). the multipolar structure of the gravitational news function at the future null infinity $\mathcal{I}^+$, the end point of future null-geodesics which escape to infinity \cite{Bondi:1962px,Penrose:1964ge}. A dynamical horizon is located inside the event horizon, which marks the boundary of a trapped space-time region. Future null infinity $\mathcal{I}^+$ is an invariantly defined null surface where outgoing null geodesics end. 

We consider $\mathcal{H}$ and $\mathcal{I}^+$ to be foliated by 2-surfaces $S$ of spherical topology, with an intrinsic Riemannian metric $q_{ab}$. For the former, we obtain a marginally trapped surface and for the latter, we approximate it by large coordinate spheres in the numerical domain. For every cross section $\mathcal{S}$, we assign outgoing and ingoing directions. Denote the outgoing future directed null vector normal to $S$ by $n^+$, and the ingoing null normal as $n^-$ satisfying $n^+\cdot n^- = -1$.  Let $x$ be a complex null vector tangent to $S$
satisfying $x\cdot\bar{x}=1$ (the overbar denotes complex conjugation),
and $n^+\cdot x = n^-\cdot x = 0$.  

In the weak-field regime, spacetime geometry is completely described by the 
Weyl tensor $C_{abcd}$.  In particular, outgoing transverse radiation 
is described by the Weyl tensor component \cite{Newman:1961qr}
\begin{equation}
 \Psi_4 = C_{abcd}n^{-a}\bar{x}^bn^{-c}\bar{x}^d\,.
\end{equation}
$\Psi_4$ can be expanded
in spin-weighted spherical harmonics ${}_{-2}Y_{\ell,m}$ of spin weight $-2$
\cite{Goldberg:1966uu}.  Let $\Psi_4^{(\ell,m)}$ be the mode component
with $\ell\geq 2$ and $-m\leq \ell \leq m$.  The $(\ell,m)$ component
of the news function $\mathcal{N}^{(\ell,m)}$ and its polarizations $\mathcal{N}_{+,\times}$ are defined as
\cite{Bondi:1962px}
\begin{equation}
  \mathcal{N}^{(\ell,m)}(u) = \mathcal{N}^{(\ell,m)}_+ + i\mathcal{N}^{(\ell,m)}_\times = \int_{-\infty}^u\Psi_4^{(\ell,m)}\,du\,.\label{news}
\end{equation}
The outgoing energy flux is related to the integral of $|\mathcal{N}|^2$ over 
all angles.  In a numerical spacetime it is in
principle possible to extract $\Psi_4$ going out all the way to
$\mathcal{I}^+$ \cite{Babiuc:2008qy}
to reduce systematic errors.  We shall however follow the common approach of
calculating $\Psi_4$ on a sphere at a finite radial coordinate $r$ and
the integral in the previous equation is over time instead of the
retarded time coordinate $u$. The lower limit in the integral is not
$-\infty$ but the earliest time available in the simulation.  The news
function is then a function of time at a fixed value of $r$, starting
from the earliest time available in the simulation. A further time
integration of $\mathcal{N}$ yields the gravitational wave strain.  

On the black hole, the basic object here is a marginally
outer trapped surface (MOTS), again denoted by  $\mathcal{S}$. This is a closed space-like
2-surface with vanishing outgoing expansion $\Theta_{+}$: 
\begin{equation}
  \Theta_{+} = q^{ab}\nabla_an^+_b = 0\,.
\end{equation}
Its scalar curvature is denoted by $\tilde{\mathcal{R}}$, the two-metric by $\tilde{q}_{ab}$, and its extrinsic curvature by $\tilde{K}_{ab}$.
The shear of the dynamical horizon is denoted by $\sigma = x^a \bar{x}^b \nabla_{a} n^+_b$. 

At $\mathcal{H}$, there are two sets of moments (the mass and angular momentum multipoles:  $\mathcal{M}_{lm}$, $\mathcal{J}_{lm}$), although defined on the foliations of the three-dimensional dynamical horizon, can be used to reconstruct the horizon geometry in a gauge invariant manner. These moments were first defined for isolated horizons \cite{Ashtekar:2004gp} and extended to axisymmetric \cite{Schnetter:2006yt} and non-axisymmetric dynamical horizons \cite{Ashtekar:2013qta}. These multipole moments can be used to study the intrinsic geometry of dynamical horizons. They have been  used in predictions of the anti-kick in binary black hole mergers \cite{Rezzolla:2010df}, in the study of the no-hair conjecture in general astrophysical environments \cite{Gurlebeck:2015xpa}, and for studying tidal deformations of black holes \cite{Cabero:2014nza, Gurlebeck:2015xpa, prasad2021tidal}. In a more recent work \cite{prasad2021tidal}, the axisymmetric tidal deformations of dynamical horizons were studied by using the source multipole moments of the dynamical horizon in binary black hole scenarios.

We now describe how to compute these moments briefly. First, a preferred coordinate system on the leaves $S$ of the dynamical horizon is constructed using an appropriately defined axial vector field $\varphi^a$  (analogous to the axial Killing vector field on an isolated Kerr horizon). We use use the method of Killing transport to compute the axial field. Once the axial field is defined, an invariant coordinate $\zeta$ (analogous to the polar coordinate variable $\cos{(\theta)}$ of the Boyer-Lindquist coordinates) can be defined. Further details can be found in \cite{Ashtekar:2004gp, Schnetter:2006yt}. Using these, one can define the mass multipole moments as:
\begin{equation}
 \mathcal{M}_{\ell m} = \frac{M_{\mathcal{S}}R_{\mathcal{S}}^l}{8\pi} \sqrt{(2l+1) \dfrac{(\ell-m)!}{(\ell+m)!}} \oint_S \Tilde{\mathcal{R}} P^m _n(\zeta) e^{-im\phi} d^2S\,, \label{mass_mult_lm}
\end{equation}
Here $\phi$ is an affine coordinate on the integral curves of the vector field $\varphi^a$, $P^m _n$ are the associated Legendre polynomials corresponding to the eigen-functions of the Laplacian on $\mathcal{S}$, and $P^{\prime m} _n $ their derivatives. In this notation, $\mathcal{M}_{00}$ is the mass $M_{\mathcal{S}}$ of the slices $\mathcal{S}$ of the dynamical horizon, and $\mathcal{J}_{10}$ is its angular momentum $J_{\mathcal{S}}$. $n!$ denotes the factorial of an integer $n$. These moments are defined for all positive $\ell$, and for each $\ell$, the azimuthal mode number $m$ takes integer values ranging from $(-\ell, \ell)$. It is to be noted that modes for which $m\neq0$ are complex in general and we denote their strength as absolute magnitude or a quadratic sum $\vert \mathcal{M}_{\ell m} \vert = \sqrt{Re(\mathcal{M}_{\ell m})^2 + Im(\mathcal{M}_{\ell m})^2}$.  

We compute and study the evolution of the mass dipole ($\ell=1$) and the quadrupole ($\ell=2$) moments, $\vert m \vert \leq \ell$ moments of the individual dynamical horizons of the black holes during the inspiral phase, and that of the common horizon in the post-merger phase. We also study their relationship with the dynamics of the binary black hole system and the gravitational news function at $\mathcal{I}^+$. 

\noindent \emph{The numerical simulations:} \label{sec:nr} Our numerical simulations are 
performed using the publicly available Einstein Toolkit framework 
\cite{Loffler:2011ay, EinsteinToolkit:web}. The initial data is generated 
based on the puncture approach \cite{PhysRevLett.78.3606,Ansorg:2004ds}, 
which is then evolved through  BSSNOK formulation
\cite{Alcubierre:2000xu, Alcubierre:2002kk, Brown:2008sb} using the
$1+\log$ slicing and $\Gamma$-driver shift conditions.  Gravitational 
waveforms are extracted \cite{Baker:2002qf} on coordinate spheres at 
various radii between $100M$ to $500M$.  
The computational grid set-up is based on the multipatch approach using 
Llama \cite{PhysRevD.83.044045} and Carpet modules, along with adaptive 
mesh refinement (AMR). The various horizons are located using the method 
described in \cite{Thornburg:1995cp, Thornburg:2003sf}. 
General quasi-local physical quantities are computed on the horizons following 
\cite{Dreyer:2002mx, Schnetter:2006yt}. The framework to compute the generalized multipole moments in Eq.~\eqref{mass_mult_lm} does not exist in the QuasiLocalMeasures thorn of the Einstein Toolkit. They were computed in post-processing involving a pythonic script using numerical relativity data. 

We consider non-spinning binary black hole systems with varying mass-ratio 
$q=M_2/M_1$, where $M_{1,2}$ are the component horizon masses ($M_1\geq M_2$). 
We use the GW150914 parameter file available from \cite{wardell_barry_2016_155394} as
our template. For each of the simulations, as input parameters we provide initial
separation between the two punctures $D$, mass ratio $q$ and the radial and azimuthal linear momenta $p_r$, $p_{\phi}$ respectively, while keeping the total physical horizon masses $M=M_1+M_2=1$ fixed in our units. Parameters are listed in table \ref{tab:ic_qc0}. We compute the corresponding initial locations, the $x$, $y$, $z$ components of linear momentum for both black holes, and grid refinement levels, etc., before generating the initial data and evolving it. We chose 2 non-spinning cases with mass-ratios $q=0.6, 0.7$ for the purposes of this study, based on the initial parameters listed in \cite{Healy:2014yta,PhysRevD.95.024037}. For computing the quasi-local quantities, we use a grid of size ($36, 74$) on each of the dynamical horizons. This grid resolution allows us to safely study multipole moments of upto $\ell=2$. We also carry out a convergence test of the horizon data by running the simulation at three different grid resolutions of the horizon. Our simulations agree very well with the catalog simulations \cite{RITcatalog:web}, with merger time discrepancies less than a few percent.  The results presented here are general features that were seen across the simulations $q=1, 0.6$, and $0.7$ described in Table~\ref{tab:ic_qc0}. We present the results using $q=0.6$. The outer common horizon of the configuration $q=0.6$ appears at at $t=1656.045M$, which we designate as merger time. When the common horizon is found, it has an areal radius of $R_{c} = 1.708$. 3D visualizations were performed using \emph{VisIt} \cite{VisIt}.
\begin{table}
\begin{center}
\begin{tabular}{|p{2cm}|p{2cm}|p{2cm}|p{2cm}|}
 \hline
 $q $ & $D/M$ & $p_r/M$& $p_{\phi}/M$\\
\hline
1   &   9.5332  &  0            &   0.09932\\
0.6 &   11.5    &  -5.46e-04    &   0.08206\\
0.7 &   12.0    &  -5.07e-04    &   0.08246\\
 \hline
\end{tabular}
\caption{Initial parameters for non-spinning binary black holes with quasi-circular orbits. $q=M_2/M_1$ is mass ratio, $D$ is the initial separation between the two holes, $p_r$ and $p_{\phi}$ are the linear momenta in the radial and azimuthal directions respectively.}
\label{tab:ic_qc0}
\end{center}
\end{table}
\begin{figure*}[htp]
\includegraphics[width=2\columnwidth]{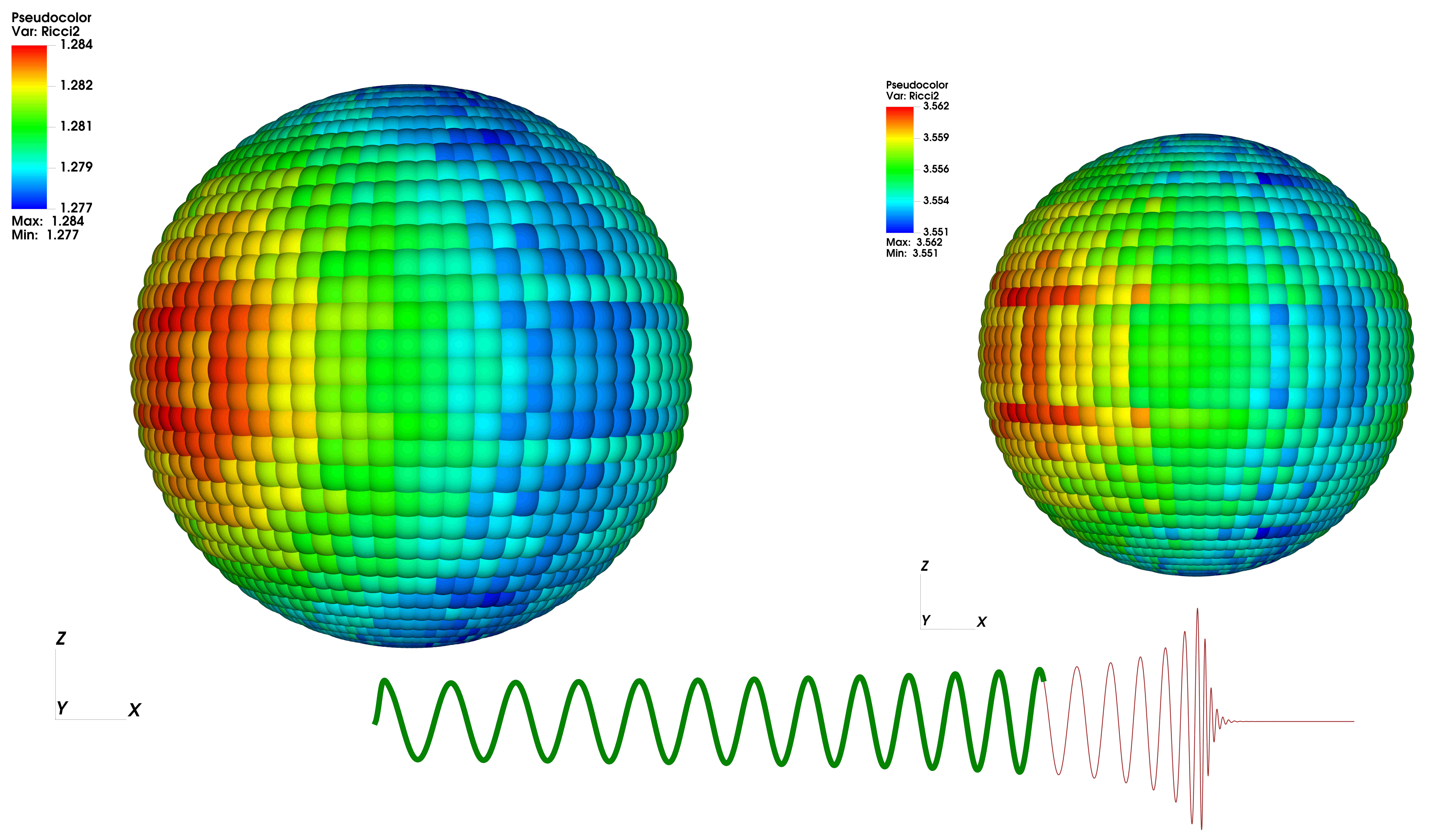}
\caption{The deformation of the dynamical horizons of the black holes can be directly visualized in terms of the 2-Ricci scalar $\Tilde{\mathcal{R}}$ of their two-dimensional slices $\mathcal{S}$. Here the 2-Ricci scalars of $\mathcal{S}$ are visualized for the $q=0.6$ system when the black holes are at a separation of $d\approx 7.75M$ (about 67.4\% of the initial separation), approximately 5 orbits after the start of the simulation, as shown by the thick green line in the waveform plot below the figure. The total number of orbits before the merger is around $9$ and the corresponding complete waveform cycles in the simulation are shown by the thin red line. The more massive black hole $BH1$ is on the left. The values of the Ricci scalar are shown on the color bars to the left of each black hole. A movie for $q=0.6$ can be viewed \href{https://drive.google.com/file/d/1HDKOEHD8LCO9CW--mvq12MaTOvV5_cvg/view?usp=sharing}{\color{blue} here} \cite{td_movie:2021}. This movie shows that the deformation of the horizon geometry is mutual for both the horizons and the Ricci scalar distribution patterns face each other at all points on the orbit. The dominant quadrupolar structure can be seen, which is reflected in the the numerical values of the strengths of the multipole moments in Fig.~\ref{fig:Mlm_time}.}
\label{fig:R2_bh12_q0p6}
\end{figure*}
\begin{figure}[htp]
\includegraphics[width=\columnwidth]{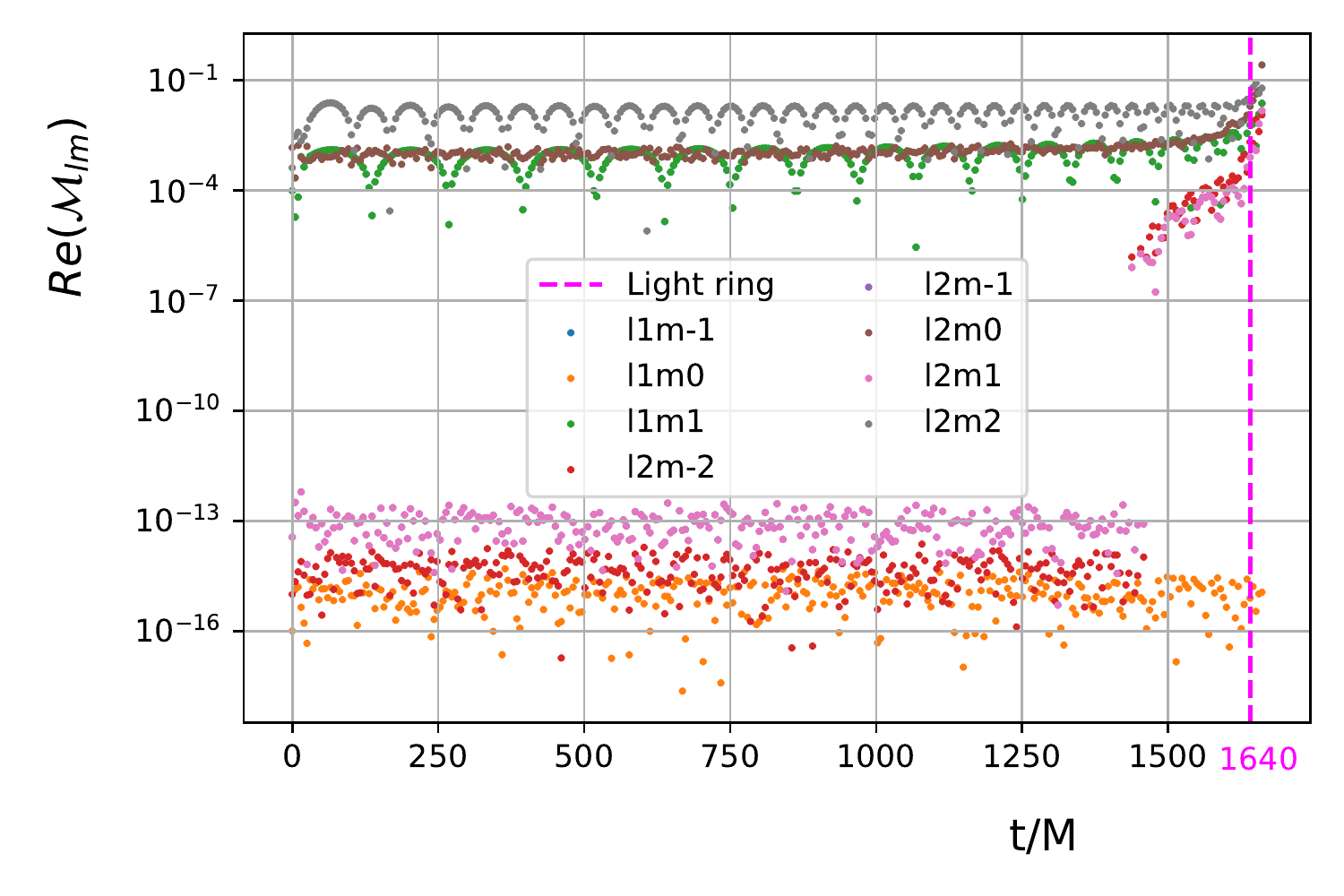}
\caption{The time evolution of the real part of the multipole moments $\ell=1, \ell=2$ $\vert m \vert < l$. Here, the values below $10^{-8}$ are below machine precision. The time of crossing of the light ring of the system is denoted as a dotted line in magenta.}
\label{fig:Mlm_time}
\end{figure}
\newline
\noindent \emph{Results: Inspiral.} We first discuss the relative strengths of the various multipole moments $\vert \mathcal{M}_{lm}\vert$. For both the black holes of all the simulations, amongst all the moments at $\ell=1, 2$ multipolar order, the multipole moment $\mathcal{M}_{22}$ was found to have the largest and monotonically increasing amplitude, followed by $\mathcal{M}_{20}$ and $\mathcal{M}_{1\pm 1}$ (note that for an isolated Kerr horizon, apart from its mass, only $\ell=2$, $m=0$ mass moment is non-zero at the quadrupolar order). This can be seen in the Figs.~\ref{fig:R2_bh12_q0p6},\ref{fig:Mlm_time} and the \href{https://drive.google.com/file/d/1HDKOEHD8LCO9CW--mvq12MaTOvV5_cvg/view?usp=sharing}{\color{blue} movie} \cite{td_movie:2021}, where the evolution of the 2D-Ricci scalars of the dynamical horizons in the inspiral phase have been visualized. In Fig.~\ref{fig:R2_bh12_q0p6}, a snapshot of the movie is presented. The multipolar deformations of the horizon geometries are mutual and are dependent on the location of the black holes in the binary system. The quadrupolar i.e. $\ell=2, m=2$ pattern can be clearly seen in the movie.
\begin{figure}[htp]
\includegraphics[width=0.9\columnwidth]{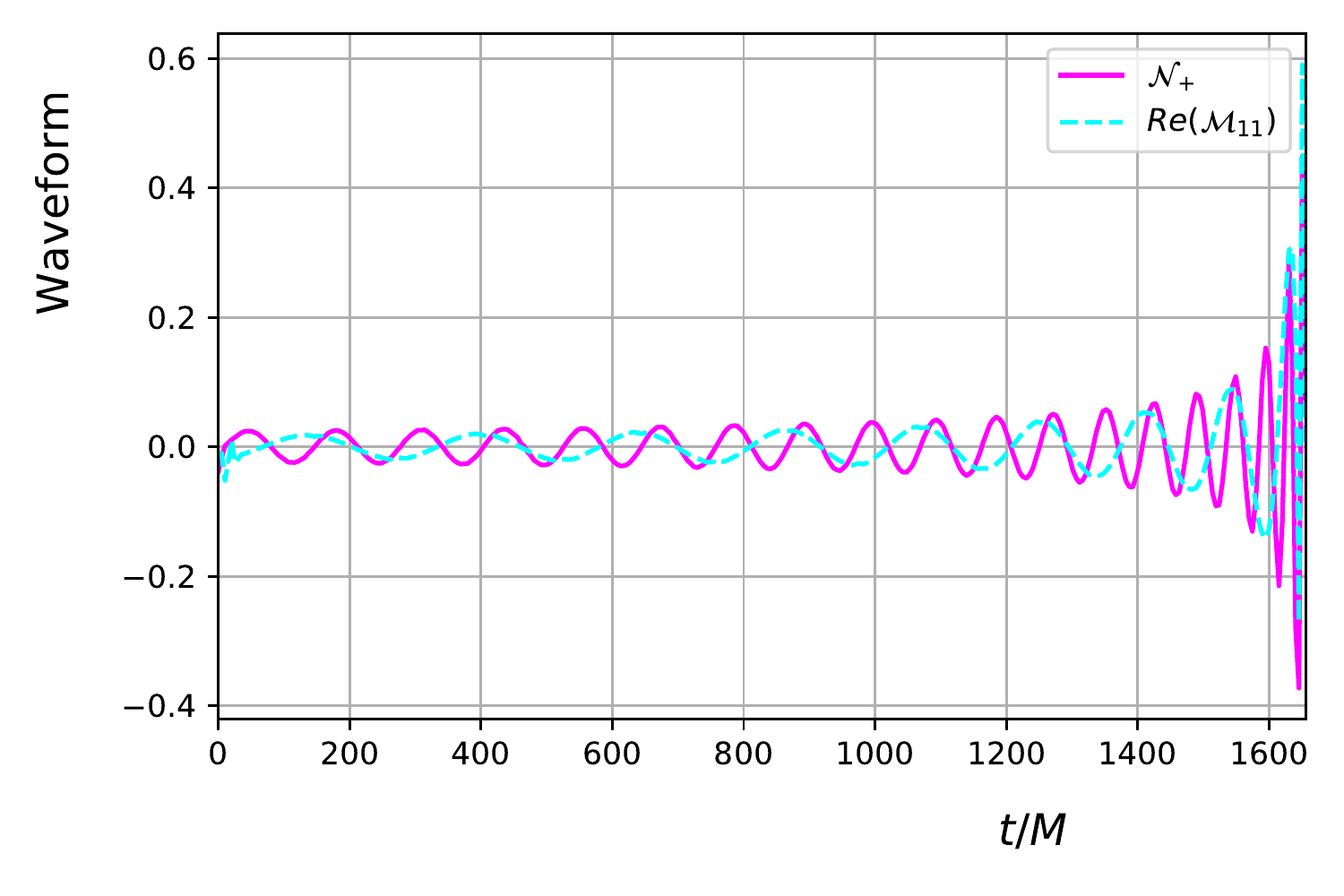}
\includegraphics[width=0.9\columnwidth]{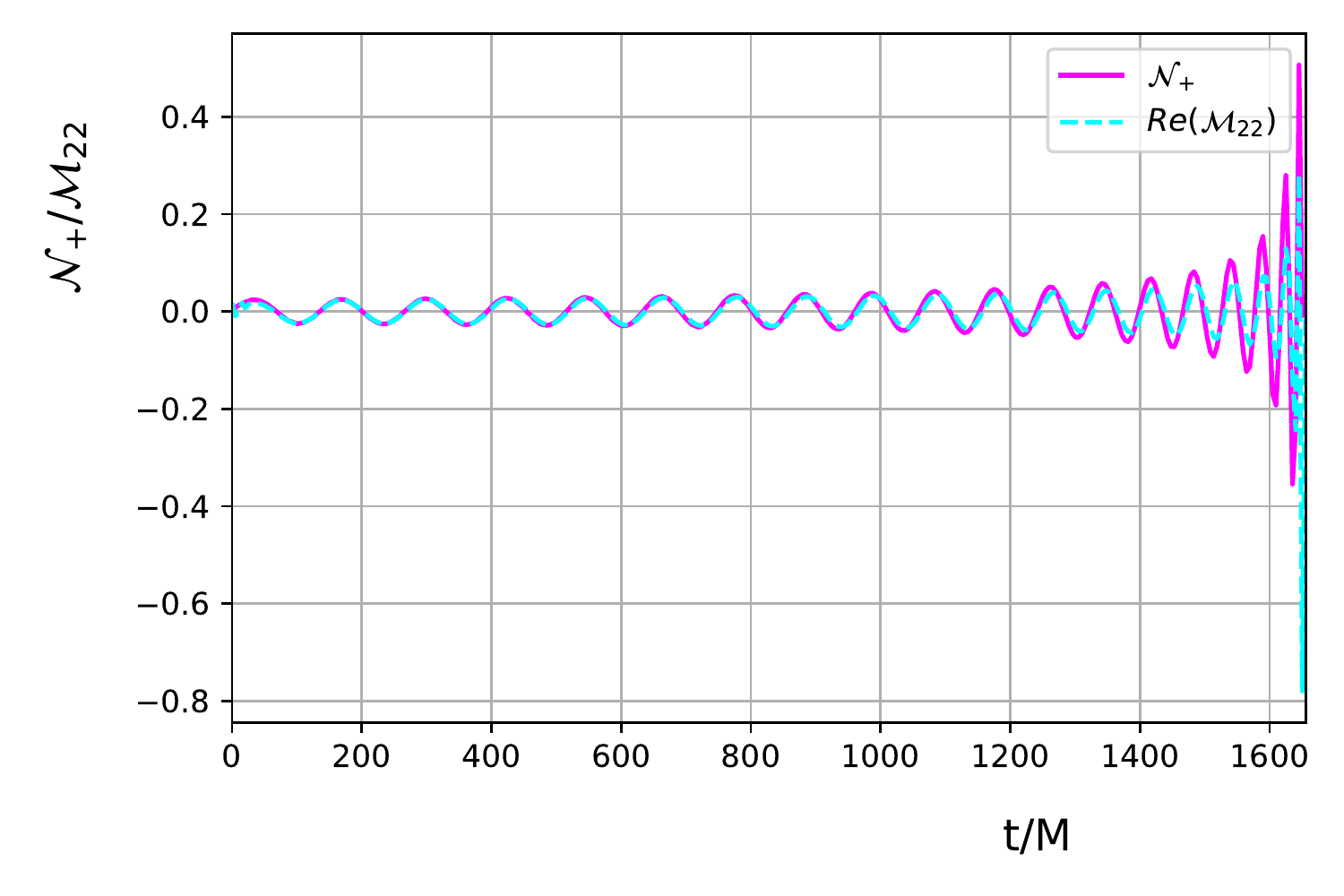}
\caption{The time derivative of the multipole moments $\mathcal{M}_{11}$ (top) and $\mathcal{M}_{22}$ (bottom) aligned suitable in phase and time with the news function of the gravitational radiation recorded at $r=100M$. Note that the phasing of the $\mathcal{M}_{11}$ is consistent with the orbital phasing of the system and that of $\mathcal{M}_{22}$ with the gravitational waveform.}
\label{fig:M11dot_Ml22dot_news}
\end{figure}
The non-axisymmetric multipole moments (i.e., $m\neq 0$) of the individual dynamical horizons of the black holes are oscillatory in nature. It was found that the multipole moments $\mathcal{M}_{22}$ of the two black holes were in-phase with each other while the moments $\mathcal{M}_{1\pm1}$ differed by a phase of $\pi$ radians.
Secondly, we found the multipole moments of dynamical horizons of the two black holes to be strongly correlated with each other. Furthermore, the dominant multipole moment $\mathcal{M}_{22}$ of the dynamical horizons is found to be strongly correlated with the dominant $(\ell=2, m=2)$ mode of the gravitational wave strain extracted at a very large distance $r=100M$ from the system. The movie \cite{td_movie:2021} aids in the visualization of some of these results. Thus, the source multipole moments are strongly correlated with the multipole moments of the gravitational field at null infinity \cite{Bondi:1962px, Sachs:1962wk, Newman:1962, janais:1965}. In Fig.~\ref{fig:M11dot_Ml22dot_news}, we plot the time derivative of the multipole moments $\mathcal{M}_{11}$ and $\mathcal{M}_{22}$ vs. the news function of the gravitational waves emitted from the system (Eq.~\ref{news}), suitably normalised and aligned in time and phase. The timeshift was found to be $101.3M$, approximately consistent with the light travel time corresponding to the extraction radius for the news. It was found that the quadrupole mass moment $\mathcal{M}_{22}$ encodes accurate information about the phasing of the gravitational waveform from the system whereas the dipole moment $\mathcal{M}_{1\pm 1}$ reflects the orbital phasing of the system. Thus instead of the wavefrom received at $\mathcal{I}^+$, these multipole moments can therefore be used to extract information of the binary black hole system like their masses, velocities, orbital angular momentum, etc. To demonstrate this, using the multipole moments and a standard least squares figure of merit, the parameters of the binary system could be estimated using the multipole moment $\mathcal{M}_{22}$. For this, we used a template bank of gravitational wave strain in the mass-ratio, chirp-mass parameter space constructed using the well known phenomenological waveform model IMRPHenomPv2. The parameters of the binary system could be estimated quite accurately (with an error of $0.12\%$ in the mass ratio and $0.01\%$ in the chirp-mass of the binary system).  
\begin{figure*}[htp]
\includegraphics[width=0.66\columnwidth]{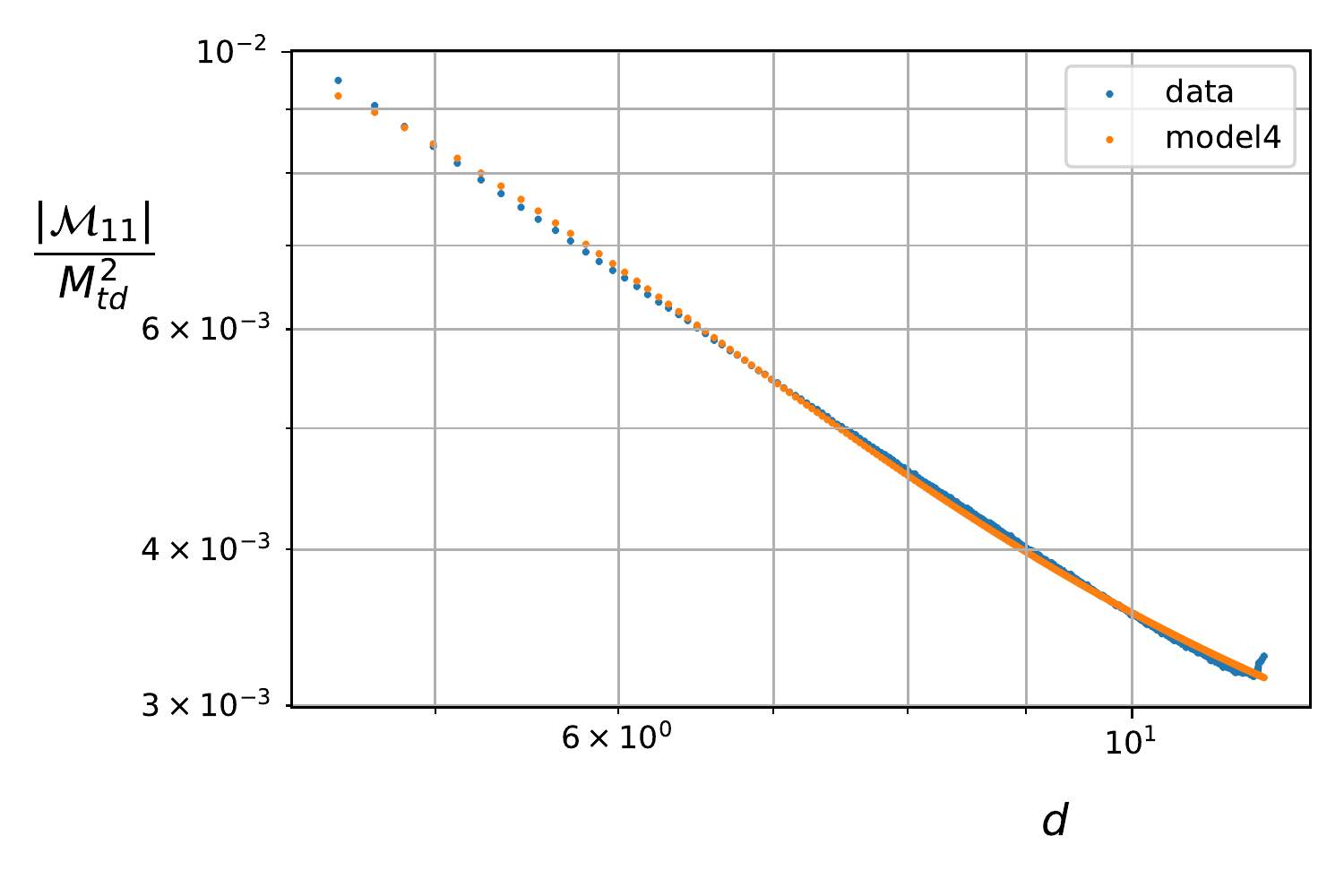}
\includegraphics[width=0.66\columnwidth]{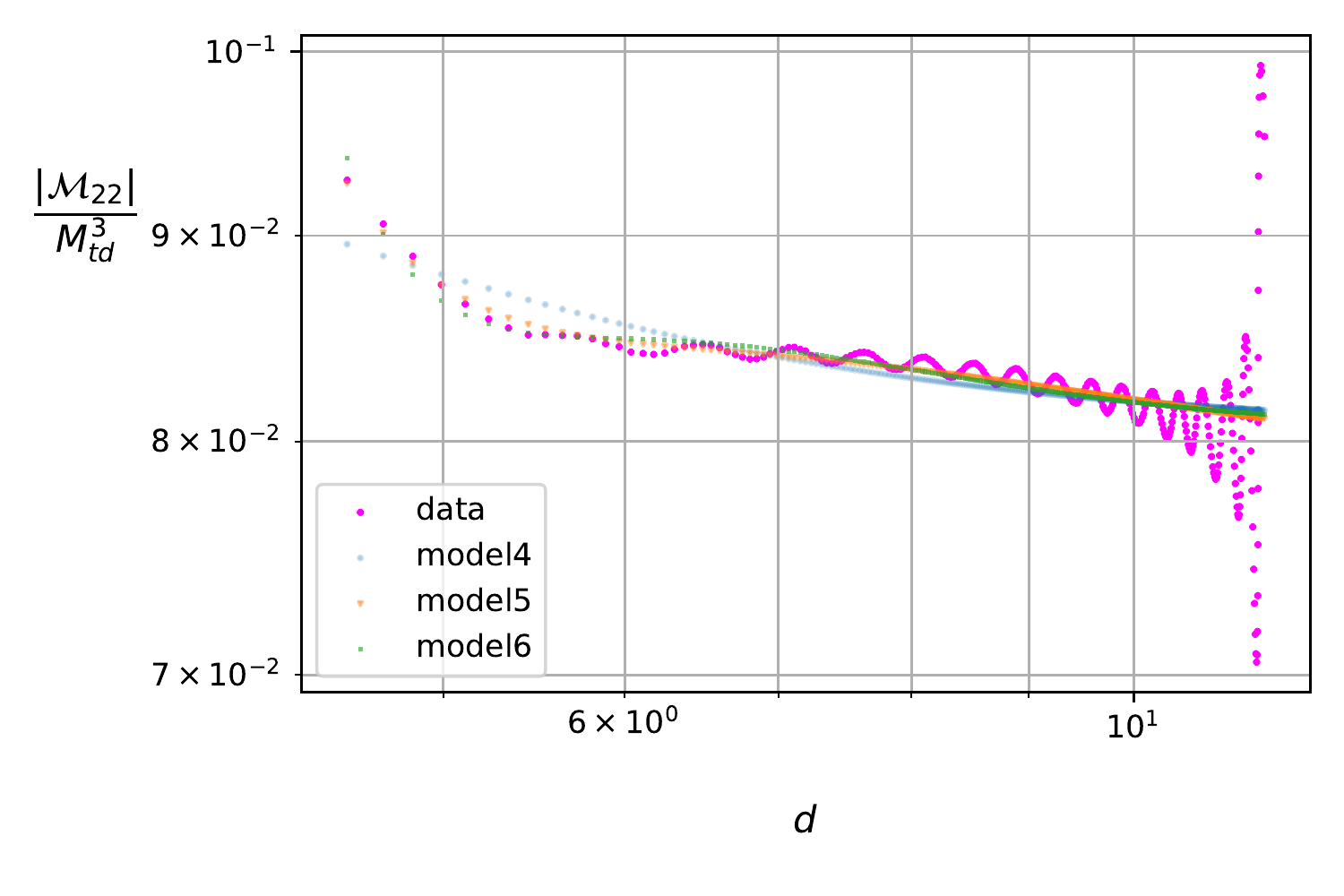}
\includegraphics[width=0.66\columnwidth]{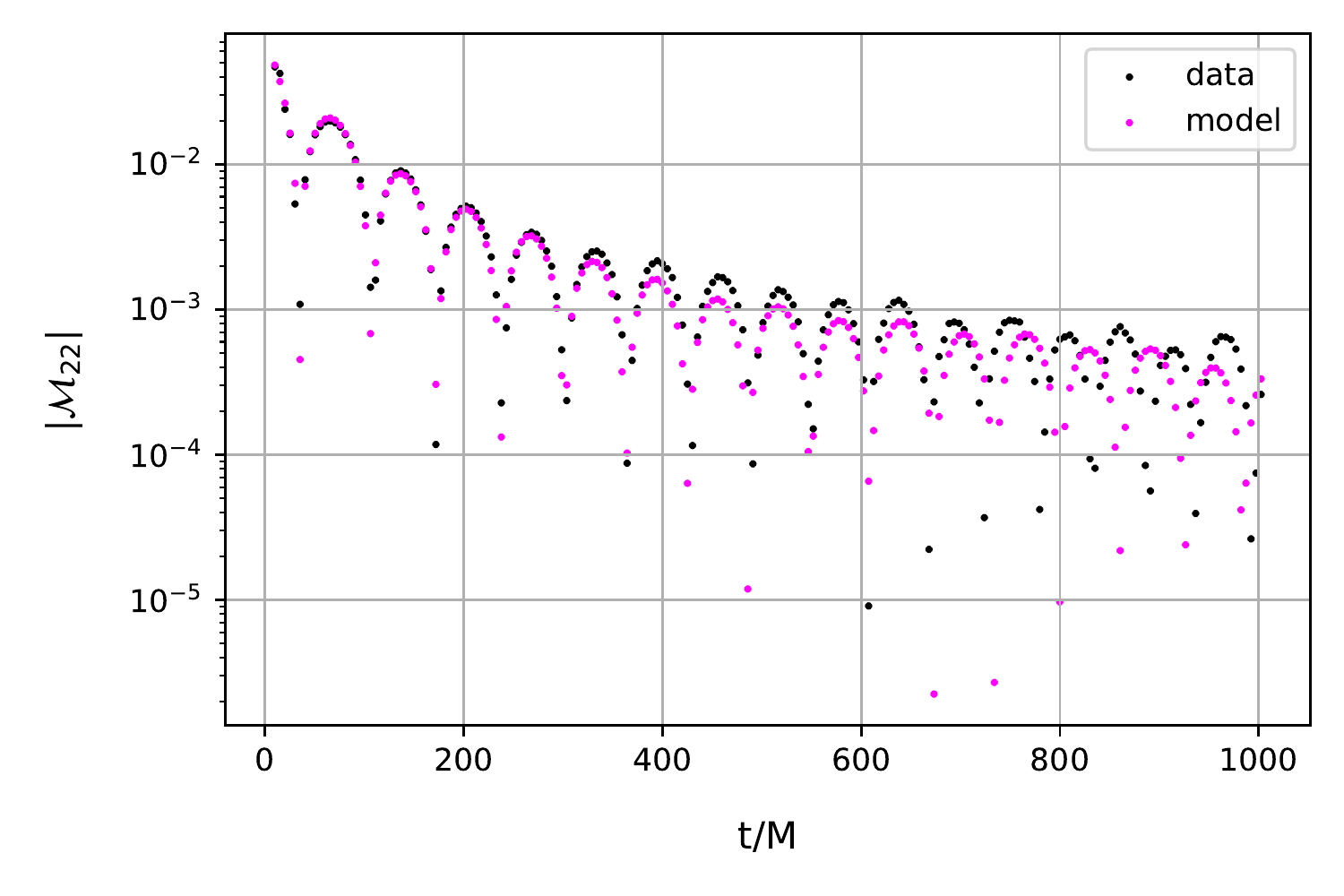}
\caption{Left : The fit of the $\ell=1, m=\pm 1$ multipole moments to the relation in Eq.~\ref{exp}, with two terms at the third and fourth order in the separation $d$, i.e. $1/d^3$ and $1/d^4$. Center:  the fit of the $\ell=2, m=\pm 2$ multipole moments to the relation in Eq.~\ref{exp}, with terms upto fourth, fifth, and sixth order in $d$. Right: the isolated oscillations of the overall amplitude of the multipole moment $\vert \mathcal{M}_{22} \vert$ seen in the figure at the center. }
\label{fig:Mlm_dist_osc}
\end{figure*}
The evolution of the strengths of the multipole moments were also found to display a generic behaviour. By means of maximizing a least-squares figure of merit, we found that the evolution of the multipole moments of both the black holes, and across the two simulations, can be described by a generic tidal expansion of the form: 
\begin{equation}
\dfrac{\delta \lvert \mathcal{M}_{lm} \rvert}{M_{\mathcal{H}}^{l+1}} = \sum_{i=3} ^{\infty} \dfrac{a_i}{d^i} \label{exp}
\end{equation}
where $M_{\mathcal{H}}$ is the mass of the black hole that is being discussed, and $d$ is a measure of distance of separation between the holes. In particular, the dipole moment was found to be well described by the above expansion that includes terms upto the order of $1/d^4$ and the quadrupole moment required terms upto the order $1/d^6$ (see Fig.~\ref{fig:Mlm_dist_osc}). These are consistent with the results of \cite{prasad2021tidal}. These moments can therefore be used to study the tidal deformability of black holes and compute its corresponding Love numbers in a manner described there. 
\begin{figure}[htp]
\includegraphics[width=0.9\columnwidth]{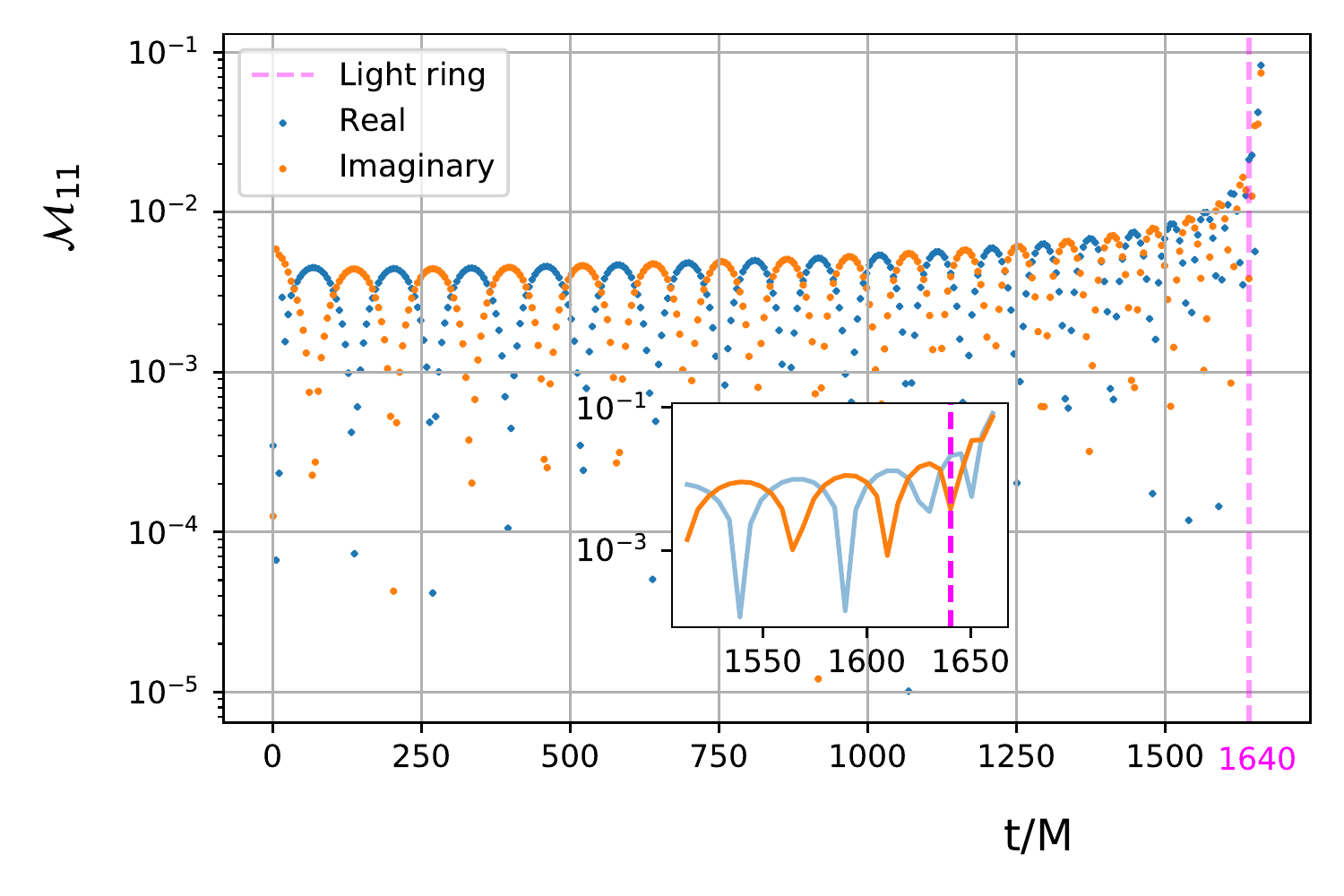}
\includegraphics[width=0.9\columnwidth]{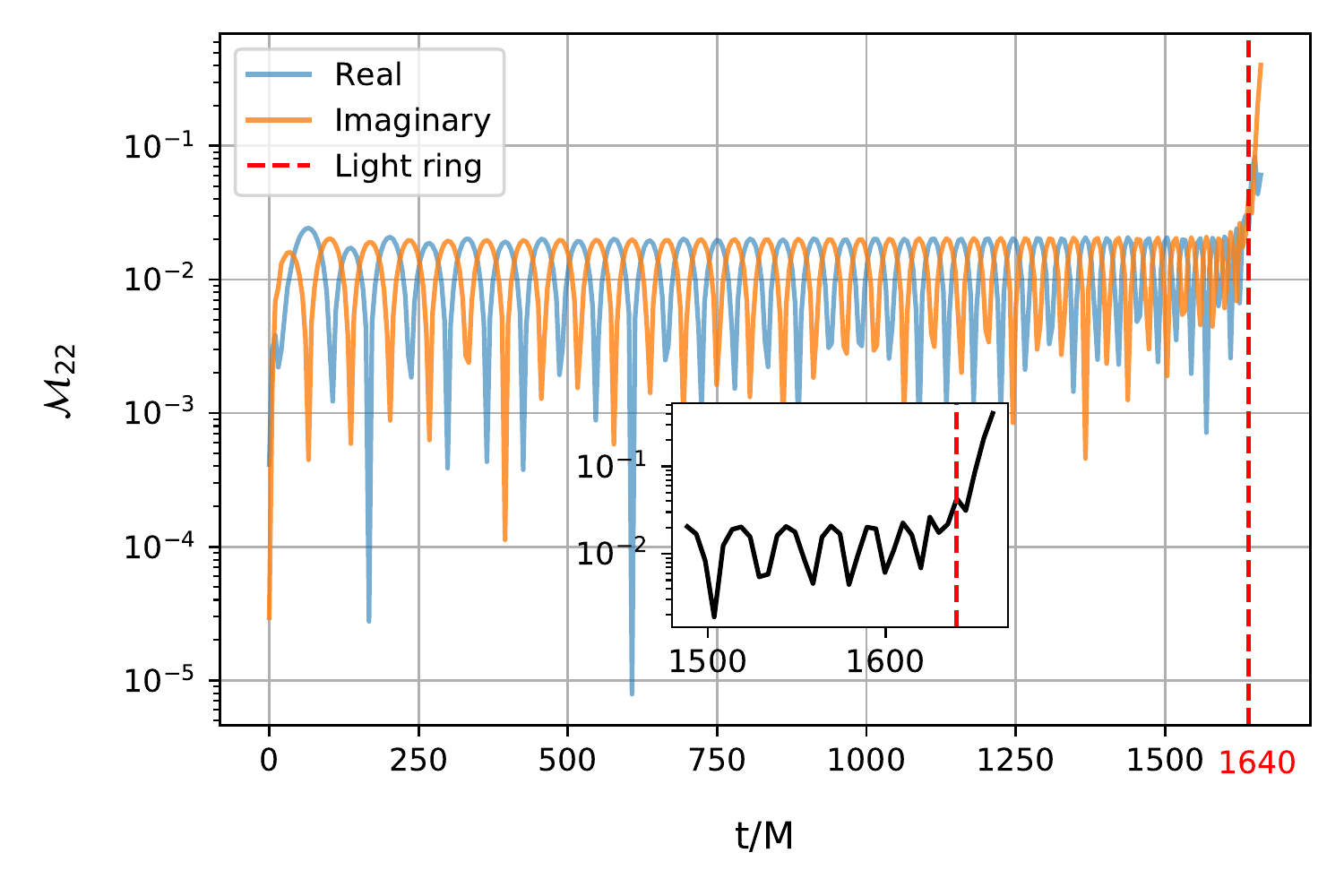}
\caption{The time evolution of some of the $\ell=1 m =1$ (top), $\ell=2, m=2$ (bottom) multipole moments. The $\ell=1, m= \pm$ moments are identical. The red line shows the temporal location of the light ring $r\approx2.856$. The close-up of these plots show that the location where the change in the growth pattern of the multipole moments occurs are consistent with the epoch of light ring crossing by the system.}
\label{fig:M11_Ml22_LCO}
\end{figure}
Apart from the real and imaginary parts of the moment, its magnitude also displays oscillatory behaviour as shown in Fig.~\ref{fig:Mlm_dist_osc}. These oscillations are decaying with time, and exist in the multipole moments of both the dynamical horizons. Isolating these oscillations in the data up to $t=1000M$, and we found that they can be described by a superposition of power-law damped sinusoids of the form:
\begin{align}
    \dfrac{\vert\mathcal{M}_{22} \vert}{M_{\mathcal{H}}^{3}} = \sum_i A_i t^{(-\gamma_i)} \sin(\omega_i t + \phi_i)
\end{align}
with power law indices $\gamma = 1.47$ and $2.29$ respectively. The time periods of oscillations of these modes were found to be at $T_1 = 161.56M$ and $T_2 = 124.71M$ respectively. It is worth noting that the latter is close to half the average orbital time period in the domain considered. These values are expected to be dependent on the mass-ratio the system.
As this feature was observed in both the simulations $q=0.6, 0.7$, we are led to speculate if these correspond to dynamic tides akin to the $w-$ mode oscillations of a star, quasi-normal modes of the tidally coupled dynamical horizons excited in the inspiral phase, or are mere numerical artefacts.

\noindent \emph{Results: Plunge.} The multipole moments were found to display two distinct behaviours in the pre-merger phase Fig.~\ref{fig:M11_Ml22_LCO}. While the majority of the portion of the evolution of these moments reflected the adiabatic, quasi-circular dynamics of the system, their behaviour change to a steeper increase in strength with the seizure of oscillations closer to the merger. We found that this epoch is very close to the time of crossing of the light ring of the binary black hole system. The light ring of the system is defined as the last circular (unstable) photon orbit. We estimate the light ring radius of the system using the adiabatic re-summed 1PN Hamiltonian \cite{buonanno2000}. It was found that the sharp change in behaviour occurs when the black holes cross the light ring of the system. 

Some of the multipole moments which were zero (below machine precision) for the majority of the inspiral phase, were found to gain in magnitude with oscillatory behaviour as the black holes approached each other.  Thus, during these last moments of the merger of the individual horizons of the black holes and before the common horizon appears, the various multipole moments of the horizons grow to comparable strengths as the dynamical horizons strongly deform under each other's influences. 

A distinctive feature was observed for $q=1$ case. The $\ell=2, m=2$ moment, although the strongest, does not monotonically increase in strength closer to the merger. This may be due to the enhanced symmetry of the equal mass binary system.
 
Owing to these and the previous results in \cite{shear-news2020}, we can conclude that the multipole moments of a dynamical horizon is strongly correlated with the dynamics of the system. It thus is also strongly correlated with the shear of its outgoing null normal, and the outgoing gravitational radiation emitted from the system.

\noindent \emph{Results: Post-merger.} 
\begin{figure*}[htp]
\includegraphics[width=0.9\columnwidth]{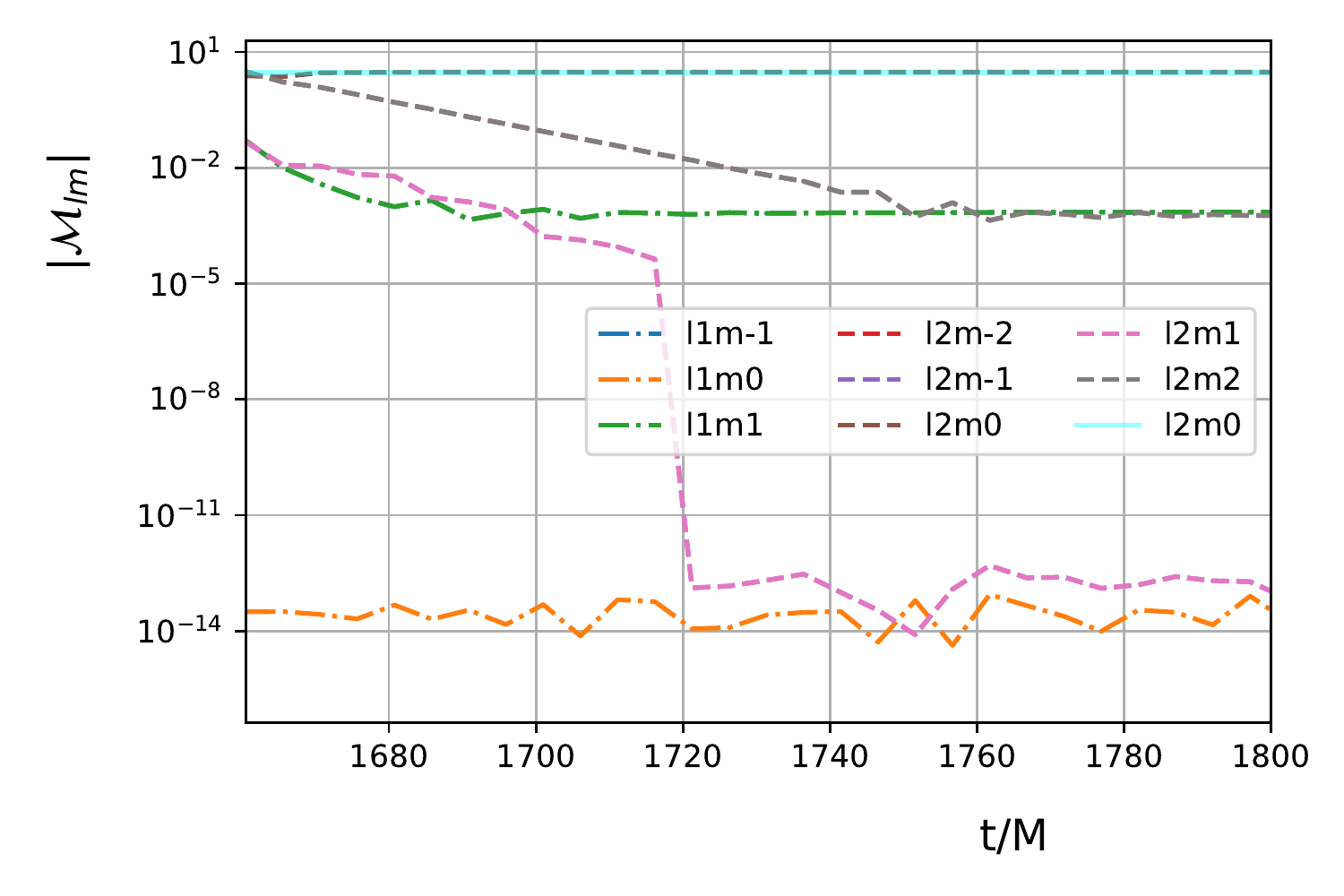}
\includegraphics[width=0.9\columnwidth]{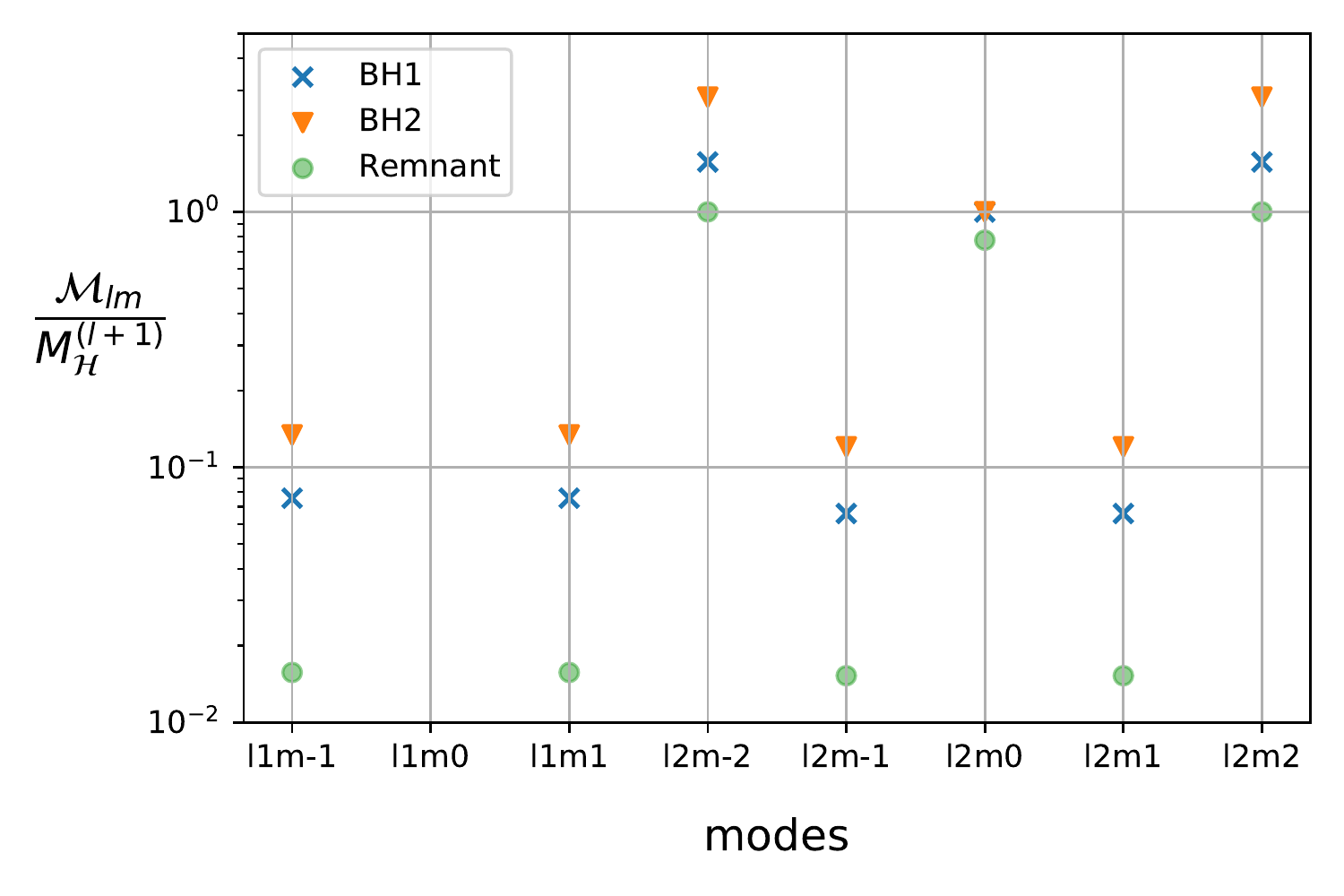}
\caption{Top: The time evolution the $\ell=1,2$ (top) multipole moments of the outer common horizon. Here, their strengths (absolute magnitudes) are plotted against time. Values $< 10^{-13}$ are below numerical accuracy. The theoretical value of the moment $\mathcal{M}_{20}$ of an unperturbed black hole of the same mass is plotted in cyan. Bottom: the strengths of the multipole moments of the individual horizons of the black holes moments before the formation of the common outer horizon of the system, and that of the common horizon just when it is formed. The moment $\ell=1, m=0$ is below machine precision as expected for all the three horizons.}
\label{fig:ml_pm}
\end{figure*}
Extending the correlations of the multipole moments of the dynamical horizons with the outgoing gravitational waves further, we analyze the multipole moments of the outer common horizon in the post-merger phase, which is formed late into the inspiral phase, immediately after it is found. 

As the two black holes cross the light ring of the system, common envelopes surrounding the individual black hole horizons form. The outer common horizon is a dynamical horizon of the remnant that settles down to the Kerr isolated horizon, which is spinning and highly distorted when formed. Thus, one expects the multipole moments of the common horizon formed to be different from that of an isolated Kerr horizon. The dynamical horizon then proceeds to equilibrium by absorbing radiation and loosing `hairs', i.e. the multipole moments are expected to decay to the corresponding isolated Kerr values. Fig.~\ref{fig:ml_pm} shows graphically that this is indeed the case. Here the time evolution of the mass multipole moments of the common horizon are plotted. The multipole moments display quasi-normal behaviour. The damping rate of the strengths of the moments $\vert \mathcal{M}_{lm} \vert$ were found to be close to the theoretical estimate of a Kerr black hole with the same mass and spin as that of the remnant. E.g, damping rate of $\mathcal{M}_{22}$ was found to be consistent with the $n=0, \ell=2, m=2$ mode with a deviation of $2.74\%$. Since the remnant black hole is spinning, the coordinate system established on the dynamical horizon can also rotate along with it and thus real part of the quasi-normal frequencies of one mode can only be estimated relative to another. Therefore the estimation of the real part of the quasi-normal frequencies requires more care and better resolution, which we will not carry out here. For an isolated Kerr horizon, the only non-zero mass multipole moment at $\ell=1,2$ order is  $\mathcal{M}_{20}$. The moment $\mathcal{M}_{20}$ approaches the value of the corresponding Kerr black hole with a final deviation of $2.73\%$ from the expected theoretical estimate. The strengths of the moments $\mathcal{M}_{1\pm1}$ and $\mathcal{M}_{22}$, which are expected to go to zero, decay to the order of $10^{-4}$. We suspect that this is due to systematic errors arising from the rotation of the coordinate system on the common horizon, and its limited grid resolution and is to be investigated further. As expected from reflection symmetry, the mass multipole moment $\mathcal{M}_{10}$ is practically zero, whereas the moment $\mathcal{M}_{2-2}$ falls below the machine precision.

What decides the deformed state of the common horizon once it is formed? The initial configuration of the parent black holes is expected to decide the deformed state of the common horizon once it is formed. Interestingly, we found that the relative strengths of the multipole moments of each of the individual horizons of the black holes just before the formation of the common horizon are similar to that of the common horizon when it is formed (more so for the horizon of the more massive black hole). This can be seen in Fig.~\ref{fig:ml_pm}. Thus the deformed state of the individual horizons just before the common horizon appears plays a role in the setting up of initial conditions of the common horizon for the post-merger dynamics. The common-horizon thus formed roughly inherits the multipolar structure of the horizon geometry of the black holes at the end of the inspiral phase and then looses them as it settles  down to equilibrium in the post-merger phase.

\noindent \emph{Conclusions:} 
In this work, the gravitational fields at the dynamical horizons in the strong-field regime and their correlation with the gravitational waves emitted from the system at future null infinity was studied in a binary black hole scenario.
The horizons of black holes in a binary environment are not isolated; their horizon geometries are dynamical and distorted due to their mutual tidal interactions, and details of the strong field dynamics of the system. In this work, these distortions were studied using the source multipole moments of the dynamical horizons. This was computed for the first time, separately using the numerical relativity data. In a binary black hole merger scenario, they are found to encode accurate information about the dynamics of the system. In particular, they display a chirp like behaviour in the inspiral phase. These moments show a distinctive behaviour towards the late inspiral phase as the dynamical horizons plunge towards each other, and is roughly consistent with the epoch of crossing of the light ring of the system. The evolution of the strengths of the source multipole moments were also found to follow a universal behaviour: they can be described by a series expansion in inverse distance of separation between the holes. 

These correlations allow one to study the strong field regime using gravitational wave observations. The correlations of the source mass multipole moments with the dynamics of the system allow the analogous, indirect interpretation that the changing source multipole moments are related to the emission of gravitational waves from the system.  We also may expect these correlations to extend to all multipolar orders, which will be studied in a future work. 

Potential applications of these results are numerous. The estimation of non-axisymmetric tidal deformability coefficients following \cite{prasad2021tidal}, can be studied. In the modelling of gravitational waveforms from binary black hole systems (e.g. in the post-Newtonian and effective one-body approach), the deformation of the dynamical horizons are not usually taken into account. As shown here, the dynamical horizons are strongly deformed in the late inspiral phase, and this can potentially have effects in the late-inspiral, plunge waveforms. 

These results urge us to state the following conjecture:  \emph{In a dynamical scenario involving binary black holes, the source multipole moments associated with charges of the dynamical horizon will be correlated with the multipolar structure of the Bondi flux of the outgoing gravitational radiation from the system received at future null infinity $\mathcal{I^+}$}

Understanding these correlations further can be useful in discerning the mass multipolar deformations of the horizons of black holes in fully dynamical scenarios using gravitational wave observations. They may also be used to test and understand further important aspects of the no-hair conjecture and probe possible non-standard structure of gravity (e.g. quantum gravitational corrections) in the strong field regime. 

The results presented here also allow for the following simple interpretation: In a binary black hole scenario, the individual dynamical horizon geometries of black holes gain a structure i.e., ``gravitational hairs" , away from their isolated Kerr geometries in the inspiral phase though the mutual tidal interactions, and loose them in post-merger dynamics of the common horizon. 

On the numerical side,  the computation of multipole moments require a choice of an axial vector field on the horizon. In this work, this has been achieved by using the method of Killing transport, which might fail very close to the merger. Accurate descriptions of the merger phase would require a better choice of the axial vector field. 

\noindent \emph{Acknowledgments:} The author thanks Sukanta Bose for valuable comments, encouragement and support, and proof-reading the manuscript. The author thanks Anshu Gupta for comments, encouragement and support. 

V.P is funded by Shyama Prasad Mukherjee fellowship (CSIR). The numerical simulations and other computations were performed on the high performance supercomputers Perseus at IUCAA.
This paper is dedicated to the loving memory of a dear senior colleague and friend, Dr. Ruchika Seth.
\section{References}
\bibliography{references.bib}
\end{document}